\documentclass[11pt,a4paper,twoside,BCOR=5mm,DIV=12]{scrartcl}
\usepackage{color}

\newcommand{\add}[1]{\textcolor{blue}{{#1}}}
\newcommand{\void}[1]{}
\renewcommand{\add}[1]{\textcolor{black}{{#1}}}

\usepackage[fleqn]{amsmath} 
\usepackage{amssymb}

\usepackage[T1]{fontenc}
\usepackage[utf8]{inputenc}

\usepackage{libertine}

\usepackage{tikz}

\DeclareSymbolFont{largesymbolstx}{OMX}{txex}{m}{n}
\DeclareMathSymbol{\intop}{\mathop}{largesymbolstx}{"52}

\makeatletter
\newcommand{\oset}[3][0ex]{%
  \mathrel{\mathop{#3}\limits^{
    \vbox to#1{\kern-2\ex@
    \hbox{$\scriptstyle#2$}\vss}}}}
\makeatother

\let\intold\int
\DeclareMathOperator*{\varrintop}{\scalerel*{\hstretch{1}{\rotatebox{5}{$\kern-.125em\displaystyle\intold\kern-.2em$}}}{\intold}}
\makeatletter
\def\varrint{\varrintop\nolimits}
\makeatother
\newcommand{\varint}[2]{\varrint_{\kern-.25em\raisebox{.02em}{$\scriptstyle\kern.000000001em #1$}}^{{\raisebox{-.02em}{$\scriptstyle\kern.025em #2$}}}}

\usepackage{inconsolata}

\usepackage{xpatch}
\usepackage{microtype}
\usepackage{blindtext}
\usepackage{lipsum}
\usepackage{siunitx}

\usepackage{scalerel}
\usepackage{graphicx}
\usepackage[explicit]{titlesec}
\usepackage{isotope}
\usepackage[square,numbers,sort&compress]{natbib}

\usepackage{booktabs}

\titleformat{\section}{\normalfont\large\bfseries}{\thesection.}{0.5em}{#1}
\setkomafont{subsection}{\normalfont\itshape}
\setkomafont{paragraph}{\normalfont\itshape}

\makeatletter
\renewcommand\maketitle[1][1]{\par
       \@tempcnta=#1\relax\ifnum\@tempcnta=1\else
         \ClassWarning{scrartcl}
                      {Optional argument of \string\maketitle\space ignored
                       at\MessageBreak
                       notitlepage-mode}\fi
       \begingroup
         \renewcommand*\thefootnote{\@fnsymbol\c@footnote}%
         \let\@oldmakefnmark\@makefnmark
         \def\@makefnmark{\rlap\@oldmakefnmark}
         \if@twocolumn
           \ifnum \col@number=\@ne
             \@maketitle
           \else
             \twocolumn[\@maketitle]%
           \fi
         \else
           \newpage
           \global\@topnum\z@   
           \@maketitle
         \fi
         \thispagestyle{plain}\@thanks
       \endgroup
       \setcounter{footnote}{0}%
       \let\thanks\relax
       \let\maketitle\relax
       \let\@maketitle\relax
       \global\let\@thanks\@empty
       \global\let\@author\@empty
       \global\let\@date\@empty
       \global\let\@title\@empty
       \global\let\@extratitle\@empty
       \global\let\@titlehead\@empty
       \global\let\@subject\@empty
       \global\let\@publishers\@empty
       \global\let\@uppertitleback\@empty
       \global\let\@lowertitleback\@empty
       \global\let\@dedication\@empty
       \global\let\author\relax
       \global\let\title\relax
       \global\let\extratitle\relax
       \global\let\titlehead\relax
       \global\let\subject\relax
       \global\let\publishers\relax
       \global\let\uppertitleback\relax
       \global\let\lowertitleback\relax
       \global\let\dedication\relax
       \global\let\date\relax
       \global\let\and\relax}

\def\@maketitle{%
    \clearpage
    \let\footnote\thanks
    \ifx\@extratitle\@empty \else
        \noindent\@extratitle \next@tpage \if@twoside \null\next@tpage \fi
    \fi
    \ifx\@titlehead\@empty \else
        \noindent\begin{minipage}[t]{\textwidth}
        \@titlehead
        \end{minipage}\par
    \fi
    \null
    \vskip 2em%
    \begin{center}%
    \ifx\@subject\@empty \else
        {\Large \@subject \par}
        \vskip 1.5em
    \fi
    {\sectfont\Large\rmfamily\bfseries \@title \par}%
    \vskip 1.5em%
    {\large
      \lineskip .5em%
      \begin{tabular}[t]{c}%
        \@author
      \end{tabular}\par}%
    \vskip 1em%
    {\large \@date \par}%
    \vskip \z@ \@plus 1em
    {\Large \@publishers \par}
    \ifx\@dedication\@empty \else
        \vskip 2em
        {\Large \@dedication \par}
    \fi
  \end{center}%
  \par
  \vskip 2em}
\makeatother

\setkomafont{caption}{\footnotesize}
\setkomafont{captionlabel}{\footnotesize\scshape}
\setcapwidth[c]{.9\textwidth}
\setcapindent{0pt}

\usepackage{fancyhdr}
\usepackage[shortlabels]{enumitem}

\usepackage{abstract}

\usepackage[hidelinks]{hyperref}
\hypersetup{pdflinkmargin=0.005cm}

\usepackage[
  bottom,
  hang,
  flushmargin,
  multiple
]{footmisc}
\renewcommand*{\thefootnote}{\arabic{footnote})}

\usepackage{braket}

\usepackage{xcolor}
\usepackage{blindtext}

\newcommand{\bargket}[1]{|\kern-.25em\ket{#1}}
\newcommand{\bargbra}[1]{\bra{#1}\kern-.25em|}
\newcommand{\vertiii}{{\vert\kern-0.2ex\vert\kern-0.2ex\vert}}

\begin{document}
\setlength{\parindent}{1em}
\thispagestyle{plain}
\title{Dynamics of interacting bosons using the Herman-Kluk semiclassical initial value representation}

\author{{\scshape Shouryya Ray}\kern.05em\thanks{Corresponding author. Electronic address: \texttt{Shouryya.Ray@mailbox.tu-dresden.de}}\kern.15em, {\scshape Paula Ostmann}, {\scshape Lena Simon}, {\scshape Frank Großmann}\\ and {\scshape Walter T. Strunz} \\[0.5em]
{\itshape Institut für Theoretische Physik, Technische Universität Dresden}\\ {\itshape D-01062 Dresden, Germany}
}
\maketitle

\vspace*{-1.8em}
\begin{abstract}
\vspace*{-.75em}
\noindent Recent experimental progress in monitoring the dynamics of ultracold gases in optical lattices necessitates a quantitative theoretical description 
for a significant number of bosons. In the present paper, we investigate if time-dependent semiclassical initial value methodology, with propagators expressed 
as integrals over phase space and using classical trajectories, is suitable to describe interacting bosons, concentrating on a single mode. 
Despite the non-linear contribution from the self-interaction, the corresponding classical dynamics allows for a largely analytical treatment of the 
semiclassical propagator. We find that application of the Herman-Kluk (HK) propagator conserves 
unitarity in the semiclassical limit, but a decay of the norm is seen for low particle numbers $n$. 
The frozen Gaussian approximation (HK with unit prefactor) is explicitly shown to violate unitarity in the present system for non-vanishing interaction strength, 
even in the semiclassical limit. Furthermore, we show by evaluating the phase space integral in steepest descent approximation, 
{that the HK propagator reproduces the exact spectrum correctly in the semiclassical limit ($n\rightarrow \infty$).} \void{In fact, 
due to a subtle cancellation of errors, the HK spectrum obtained in this manner is in full agreement with the exact quantum mechanical solution.} An error is, 
however, incurred in {next-to-next-to-leading order (small parameter $1/n$), as seen upon numerical evaluation of the integral
and confirmed analytically by considering finite $n$ corrections to the steepest descent calculations.} 
The frozen Gaussian approximation, in contrast, is only accurate to {lowest order}, and an erroneous 
{next-to-leading order} term in the energy spectrum was found analytically. Finally, as an example application, we study the dynamics of 
wave packets by computing the time evolution of the Wigner function. While the often-used truncated Wigner approximation cannot capture any 
interferences present in the exact quantum mechanical solution (known analytically), we find that the HK {approach}, despite also using classical 
information only, reproduces the salient features of the exact solution correctly.\\[.75em]
\end{abstract}

\section{Introduction}
\label{sec_intro}
Recent experimental progress has rekindled interest, from the point of view of quantum optics, in several models that originally arose in 
descriptions of solid state physics. The Bose-Hubbard model is one of the most prominent representatives in this category, having seen a surge 
in activity due to its realizability using ultracold Bose gases in optical lattices, and the implementation of numerous sophisticated experiments building 
on such technology \cite{lit_Foelling,lit_Trotzky1,lit_Trotzky2}. In terms of theoretical methodology, semiclassical \add{approaches} 
constitute \add{an important} part of the theoretical atomic, molecular and chemical physicist's toolbox. When working in the energy \add{domain}, 
Wentzel-Kramer-Brillouin (WKB) quantization and related approximations are often sufficient \cite{lit_BerryMount}. Semiclassical propagators, however, 
become indispensable if one is interested in time-dependent quantities and questions of dynamics, such as the time evolution of wave packets or other 
observables. The Herman-Kluk (HK) semiclassical initial value representation (SCIVR) {is the most widely used among the} semiclassical 
approximations to the time evolution operator \cite{lit_HermanKluk,LS77,TH91,Setal93,Mi05,Kay05,lit_GrossmannCommentsAtMolPhys,lit_FemtosecPhys}.

The aim of the present article is to leverage the methods from the latter field in order to investigate an issue of the former; 
or to put it more explicitly, we are interested in studying the application of the HK-SCIVR to the dynamics of the Bose-Hubbard model. Rather than discussing 
the full many-body problem, we restrict ourselves to the zero-dimensional (0D) case, i.e. the single-site Bose-Hubbard model, for reasons that will be 
discussed shortly.

The semiclassical treatment of the Bose-Hubbard model is not entirely without precedent. Such work has mainly been concerned with obtaining the spectrum using 
WKB and other semiclassical quantization schemes (cf. e.g. Refs. {\cite{GK07,lit_Kolovsky,lit_Itin,GGR15}}). An exception is \cite{lit_Simon1}, where WKB 
quantization was applied to the Josephson dynamics of a Bose-Einstein condensate in a double-well. The investigation most spiritually kindred to the present 
one is the recent work by Simon and Strunz \cite{lit_Simon2}, who---for the first (and only) time known to the authors---applied the HK propagator to the 
dynamics of the two- and three-site Bose-Hubbard chain. Urbina and co-workers have applied a semiclassical approximation of the propagator
to the problem of backscattering in interacting Bose-systems \cite{Urbina}.

Notwithstanding the wealth of activity in many-body quantum dynamics of cold atom systems, the 0D Bose-Hubbard model presents an attractive subject of 
investigation, because it provides an ideal test case for studying the application of the HK-SCIVR. It is on one hand involved enough, since the on-site 
interaction term of the Hamiltonian represents an anharmonicity (a term higher than quadratic), for which the HK propagator is nontrivial. On the other hand, 
the system is simple enough to permit all the necessary calculations to be carried out analytically, and hence affords a better insight into the `inner workings' 
of the HK-SCIVR itself.\footnote{By contrast, the treatment of the double- and triple-well Bose-Hubbard model using the HK-SCIVR already needs recourse to 
numerics \cite{lit_Simon2}.} This is all the more valuable, because the only other system for which analytical results are possible is the harmonic oscillator, 
by and large the most ubiquitous toy model in all of theoretical physics. The harmonic oscillator, however, is not as interesting from a Herman-Kluk perspective,
because the HK-SCIVR yields the exact time evolution operator for all Hamiltonians containing at most quadratic terms. An added advantage is that the 
zero-dimensional Bose-Hubbard model, in spite of its nontrivial features, still allows to find the exact quantum mechanical solution by elementary means, 
so that it is available as a reference point against which to compare our calculations. In summary, therefore, the choice of model system, namely the 0D 
Bose-Hubbard model, is such that it allows a great deal of the investigation to be carried out in exact and analytical manner, while still capturing nontrivial 
features of interest. As a second and more extrinsic justification, it should be kept in mind that the zero-dimensional Bose-Hubbard model is not necessarily 
just a `constructed' or `academic' curiosity: Rather, the single-site Bose-Hubbard model arises naturally when writing down an effective Hamiltonian for the 
Bose-Hubbard chain (one dimension, multiple sites) in the limit of small but finite interaction strengths.

The exposition is structured as follows: First, we recapitulate the formalism of the HK-SCIVR 
(Section \ref{sec_hkpropag}). In Section \ref{sec_hkpropag}, we then concern ourselves with its application to the problem at hand. To this end, we briefly 
review the zero-dimensional Bose-Hubbard model and its exact quantum mechanical solution, following which we finally evaluate the HK propagator for the 0D Bose-Hubbard model and derive exact analytical expressions for it in terms of an integral representation. In Sections \ref{sec_anal_results} and 
\ref{sec_numresults}, we further expound on this result by evaluating the integral analytically in \add{the steepest descent} approximation 
and numerically for more general parameter values. Last but not least, in Section \ref{sec_wavepacketdyn}, we consider an example application, viz. the dynamics of wave packets. To this end, we compute the time evolution of the Wigner function using the HK propagator and compare with the exact solution (available analytically) as well as the often-used truncated Wigner approximation (TWA). Our exposition concludes with Section \ref{sec_conclu}, in which we summarize our results and discuss 
approaches for further work.

\section{Semiclassical method}
\label{sec_hkpropag}
The task of the semiclassical propagator is to find an approximation to the time evolution operator $\hat{U}(t,t_0)$ given a Hamilton \add{operator}
$\hat{H}$. 
We assume that $\hat{H}$ has no explicit time dependence, as a result of which it follows that\footnote{Throughout this article, unless mentioned otherwise, 
we use natural units such that $\hslash = 1$.} $\hat{U}(t,t_0) = {\rm e}^{-{\rm i}(t - t_0)\hat{H}} \equiv \hat{U}(t-t_0)$. In the following, we set $t_0 = 0$, 
so that we are only interested in the quantity $\hat{U}(t) = {\rm e}^{-{\rm i}t\hat{H}}$. We shall further be using the `complex $z$ notation' of Shalashilin 
and Child (cf. e.g. Ref. \cite{lit_Shalashilin}); not only because it is more compact than the `real $qp$ notation' preferred in theoretical chemistry 
\cite{lit_FemtosecPhys}, but also because it will turn out to be more natural for describing the classical dynamics underlying second quantized Hamilton operators.

\subsection{Preliminaries and notation} Here, we introduce the necessary quantities that enter the evaluation of the HK-SCIVR, and simultaneously take 
the opportunity to fix some notation. We assume that there are $m$ degrees of freedom, with corresponding annihilation operators 
$\hat{\bf a} \equiv (\hat{a}_1,\hat{a}_2,\ldots,\hat{a}_m)$ and a normal-ordered second-quantized Hamilton \add{operator} 
$\hat{H}(\hat{\bf a},\hat{{\bf a}}^\dagger)$. 
The coherent state $\ket{{\bf z}}$ is given by\footnote{In particular, this coincides with the definition from quantum optics (cf. any standard reference 
on quantum optics, e.g. \cite{lit_Klauder,lit_quantum_optics_book}). There is a difference in phase between this convention, also called the Klauder convention, and the common 
definition of Gaussian $\ket{{\bf q},{\bf p}}$ states encountered in theoretical chemistry \cite{lit_Shalashilin}.} 
$\ket{\bf z} \equiv {\rm e}^{-\frac{1}{2}|{\bf z}|^2}{\rm e}^{\hat{{\bf a}}^\dagger{\bf z}}\ket{{\pmb 0\kern-.05em}}$ with the vacuum $\ket{\pmb 0\kern-.05em}$.
The well-known overcompleteness relation reads
\begin{equation}
\braket{{\bf z} | {\bf z}^\prime} = {\rm e}^{-\frac12(|{\bf z}|^2+|{\bf z}^\prime\kern-.05em|^2)} {\rm e}^{{\bf z}^\dagger {\bf z}^\prime} \qquad 1\kern-.3em1 = \varint{}{} \frac{{\rm d}^{2m}z}{\pi^m}\,|{\bf z}\rangle\langle{\bf z}|\;,
\end{equation}
where the integration measure \[{\rm d}^{2m} z = \prod_k{\rm d}(\operatorname{Re} z_k)\,{\rm d}(\operatorname{Im} z_k)\] is the area element of the complex plane. 
The coherent states obey the eigenvalue relation $\hat{{\bf a}}\ket{\bf z} = {\bf z}\ket{{\bf z}}$. Both of these identities can be shown by expanding 
$\ket{\bf z}$ in terms of occupation number eigenstates. The classical Hamiltonian is found as a coherent state matrix element of the form 
$H({\bf z},{\bf z}^*) = \bra{{\bf z}}\!\hat{H}(\hat{\bf a},\hat{\bf a}^\dagger)\!\ket{{\bf z}}$, which simply amounts, in view of the eigenvalue relation, to the 
substituion $\hat{\bf a} \rightarrow {\bf z}$ and likewise for the Hermitian conjugate. In complex notation, the canonical equations of motion read
\begin{equation}
{\rm i}\kern.05em\partial_t {\bf z} = \partial_{{\bf z}^*} H\;.
\label{eq_canonicaleom}
\end{equation}
The action is defined as
\begin{equation}
S\!\left[{\bf z}(t),{\bf z}^*\kern-.075em(t)\right] = \varint{0}{\kern.1em t} \!{\rm d}t^\prime L\!\left({\bf z}(t^\prime),{\bf z}^*\kern-.075em(t^\prime),t^\prime\right),
\label{eq_defS1}
\end{equation}
with the Lagrangian in turn defined as
\begin{equation}
L = \tfrac{1}{2}{\rm i}\left({\bf z}^\dagger (\partial_t\kern.05em {\bf z}) - (\partial_t\kern.05em {\bf z})^\dagger {\bf z}\right) - H\;.
\label{eq_defS2}
\end{equation}
This differs from the usual Legendre transform definition commonly found in analytical mechanics by a total derivative which compensates for the difference 
in phase convention in the definition of the coherent states $\ket{{\bf z}}$ on one hand and the Gaussian states $\ket{{\bf q},{\bf p}}$ on the other 
\cite{lit_Shalashilin}. Finally, the monodromy matrix
\begin{equation}
\underline{\bf M} = \left(\!\begin{array}{ll}
\partial_{{\bf z}(0)}\kern-.1em\otimes {\bf z}(t) & \partial_{{\bf z}^*\kern-.1em(0)}\kern-.1em\otimes {\bf z}(t) \\
\partial_{{\bf z}(0)}\kern-.1em\otimes {\bf z}^*\kern-.1em(t) & \partial_{{\bf z}^*\kern-.1em(0)}\kern-.1em\otimes {\bf z}^*\kern-.1em(t)
\end{array}\!\!\right) \equiv \left(\!\begin{array}{ll}
{\bf M}_{{\bf z}{\bf z}} & {\bf M}_{{\bf z}{\bf z}^*} \\
{\bf M}_{{\bf z}^*{\bf z}} & {\bf M}_{{\bf z}^*{\bf z}^*}
\end{array}\!\!\right)\;
\label{eq_defmonodromy}
\end{equation}
\add{is a measure of the stability of a trajectory under perturbation of the initial value.}

\subsection{The Herman-Kluk semiclassical initial value representation}
If the actions appearing in the problem are large (in conventional units compared to $\hslash$, in natural units compared to unity), then the Feynman path 
integral \add{for the propagator} can be evaluated in a stationary phase approximation. This is the starting point of most semiclassical derivations. 
The HK propagator has been considered by several researchers and derived in different ways. The original work was carried out by the eponymous 
Herman and Kluk \cite{lit_HermanKluk}, building upon and modifying the heuristic `Frozen Gaussian Wavepacket Dynamics' of Heller \cite{lit_Heller}. \add{A seminal contribution to the field is} the work of Kay \cite{lit_Kay}, who derived 
a uniform asymptotic series expansion for the propagator, of which the HK-SCIVR turns out to be the leading order contribution. 
In `complex $z$ notation', the Herman-Kluk propagator reads
\begin{equation}
\hat{U}_{\rm HK}(t) = \varint{}{} \frac{{\rm d}^{2m}z_0}{\pi^m}\, R_t\kern-.0375em\!\left({\bf z}_0^{\vphantom{*}},{\bf z}^*_0\right){\rm e}^{{\rm i}S_t\kern-.0375em({\bf z}_0^{\vphantom{*}},{\bf z}^*_0)} 
\ket{{\bf z}_t\kern-.0375em\!\left({\bf z}_0^{\vphantom{*}},{\bf z}_0^*\right)\kern-.05em}\!\bra{{\bf z}_0^{\vphantom{*}}}\;,
\label{eq_HKgen}
\end{equation}
where we have largely followed Child and Shalashilin \cite{lit_Child} in `translating' results from `real $qp$ notation' to `complex $z$ notation'.

The integral is to be taken over the whole phase space $\mathbb{C}^m$ of initial conditions ${\bf z}_0$. The trajectory 
${\bf z}(t) = {\bf z}_t\kern-.0375em({\bf z}_0^{\vphantom{*}},{\bf z}_0^*)$ now refers to the solution of Hamilton's equations of motion 
\eqref{eq_canonicaleom} with ${\bf z}(0) = {\bf z}_0$ as initial condition---the notation makes explicit the dependence on the initial value, and that it 
may not be a holomorphic function of the initial conditions. The quantity $S_t\kern-.0375em({\bf z}_0^{\vphantom{*}},{\bf z}^*_0)$ is the action evaluated 
along the classical trajectory, i.e. 
$S_t\kern-.0375em({\bf z}_0^{\vphantom{*}},{\bf z}^*_0) \equiv S[{\bf z}_t^{\vphantom{*}}\kern-.075em({\bf z}_0^{\vphantom{*}},{\bf z}_0^*),{\bf z}_t^*\kern-.075em({\bf z}_0^{\vphantom{*}},{\bf z}_0^*)]$. Finally, $R_t\kern-.0375em({\bf z}_0^{\vphantom{*}},{\bf z}^*_0)$ 
is the prefactor. This in turn comprises two separate factors of the form
\begin{equation}
R_t\kern-.0375em({\bf z}_0^{\vphantom{*}},{\bf z}^*_0) = {\rm e}^{{\rm i}\theta_t\kern-.0375em({\bf z}_0^{\vphantom{*}},{\bf z}^*_0)} R_t^{\rm \kern.05em HK}\kern-.075em({\bf z}_0^{\vphantom{*}},{\bf z}^*_0)\;.
\end{equation}
The first factor is a phase correction and is given by
\begin{equation}
\theta_t\kern-.0375em({\bf z}_0^{\vphantom{*}},{\bf z}^*_0) = \varint{0}{\kern.1em t} \!{\rm d}t^\prime\,\tfrac{1}{2}\operatorname{Tr}\kern.05em\bigl(\!\left.\partial_{{\bf z}^*}\kern-.175em\otimes \partial_{{\bf z}^{\vphantom{*}}}\kern.075em H^{\vphantom{*}}_{\vphantom{*}}\right|_{{\bf z} = {\bf z}_{t^\prime}\kern-.05em({\bf z}_0^{\vphantom{*}},{\bf z}_0^*)}\bigr)\;.
\end{equation}
It may be interpreted as a local shift of the zero-point energy \cite{lit_Shalashilin}; technically, it arises from the commutator terms required to transform 
between the symmetrically ordered Hamiltonian and the normal-ordered one \cite{lit_Child}. The second factor contains information about the stability of the 
trajectory ${\bf z}_t\kern-.0375em({\bf z}_0^{\vphantom{*}},{\bf z}_0^*)$ in the following manner:
\begin{equation}
R_t^{\rm \kern.05em HK}({\bf z}^{\vphantom{*}}_0,{\bf z}^*_0) = \sqrt{\det {\bf M}_{\bf zz}}\;.
\end{equation}
Here, ${\bf M}_{\bf zz}$ is the upper left block of the monodromy matrix \eqref{eq_defmonodromy} evaluated along the classical trajectory 
${\bf z}_t^{\vphantom{*}}\kern-.075em({\bf z}_0^{\vphantom{*}},{\bf z}_0^*)$. The square root is to be taken such that the phase is a continuous 
function of time \cite{lit_KayJChemPhys1994}.

\section{The zero-dimensional Bose-Hubbard model}
\label{sec_zerodim}
Having the necessary tools at hand, we now turn our attention to the problem at hand: the 0D Bose-Hubbard model.

\subsection{Preliminaries and notation} First, we recapitulate the 0D Bose-Hubbard model, and take this opportunity to fix some notation as well. 
One way to arrive at the 0D Bose-Hubbard model is to take the Hamiltonian of the Bose-Hubbard chain \cite{lit_JakschZoller} and restrict it to a single site, 
to wit:
\begin{equation}
\hat{H} = \omega_{\rm e}\kern.1em\hat{a}^\dagger \hat{a} + \tfrac{1}{2}U \hat{a}^\dagger \hat{a}^\dagger \hat{a}\kern.1em\hat{a}\;.
\label{eq_0DBHHamiltonian}
\end{equation}
The first term is just a harmonic oscillator with zero ground-state energy, and is solved exactly by the HK-SCIVR. The frequency of the harmonic oscillator is 
$\omega_{\rm e}$. The second term is more interesting, because it is higher order, i.e. anharmonic, so that the HK-SCIVR is not expected to be exact a priori. 
The parameter $U$ physically represents an on-site interaction energy.

The occupation number operator $\hat{n} = \hat{a}^\dagger\hat{a}$ commutes with $\hat{H}$, owing to the global $U(1)$ symmetry. The exact quantum 
mechanical solution is now elementary. The Hamiltonian can in fact be rewritten using the canonical bosonic commutation relations in terms of $\hat{n}$ to obtain
\begin{equation}
\hat{H} = \omega_{\rm e}\kern.05em\hat{n} + \tfrac{1}{2}U \hat{n}\kern.05em(\hat{n} - 1)\;.
\end{equation}
This allows to read off the eigenenergies as 
\begin{equation}
E_n = \omega_{\rm e}n + \tfrac{1}{2}Un(n-1)\;.
\label{eq_exactspectrum}
\end{equation}
By expanding in terms of occupation number eigenstates
\[
\ket{n} = \frac{(\hat{a}^\dagger)^n}{\sqrt{n!}}\ket{0}\;,
\]
one finds the exact quantum mechanical solution for the time evolution operator as\footnote{The spectrum of the 0D Bose-Hubbard model has no 
{\it essential} degeneracy. This does not, of course, preclude the occurence of {\it accidental} degeneracy, e.g.: 
For $\omega_{\rm e} = 1$ and $U = -1$, $E_0 = E_2 = 0$. This is, however, of no consequence to the expansion of the propagator in the $\{\ket{n}\}$ basis.}
\begin{equation}
\hat{U}(t) = \sum_n {\rm e}^{-{\rm i}t E_n} \ket{n}\!\bra{n}\;.
\label{eq_Uspectral}
\end{equation}

\subsection{Semiclassical limit} 

As a final preliminary before studying Herman-Kluk theory for the 0D Bose-Hubbard model, we consider the limit in which 
such a treatment is expected to be appropriate. We follow the argumentation of \cite{lit_Simon2}, adapting it to the present scenario by directly working 
with annihilation and creation operators $\hat{a},\hat{a}^\dagger$ instead of field operators. Starting from the basic Hamiltonian 
\eqref{eq_0DBHHamiltonian}, we introduce rescaled operators $\hat{\mathcal{A}},\hat{\mathcal{A}}^\dagger$ with
an average ${\overline n} = \langle \hat{a}^\dagger \hat{a} \rangle$ such that
\begin{equation}
\hat{\mathcal{A}} \equiv \frac{1}{\sqrt{\overline n}}\kern.05em\hat{a} \qquad\qquad \hat{\mathcal{A}}^\dagger \equiv \frac{1}{\sqrt{\overline n}}\kern.05em\hat{a}^\dagger \qquad\qquad \bigl[\hat{\mathcal{A}},\hat{\mathcal{A}}^\dagger\bigr] = \frac{1}{\overline n}\;
\label{eq_rescaledops}
\end{equation}
to obtain
\begin{equation}
\hat{H} = {\overline n}\left(\omega_{\rm e}\kern.05em \hat{\mathcal{A}}^\dagger \kern-.15em\hat{\mathcal{A}} + \tfrac{1}{2}U\kern.1em{\overline n}\kern.2em\hat{\mathcal{A}}^\dagger\kern-.15em\hat{\mathcal{A}}^\dagger\kern-.15em\hat{\mathcal{A}}\kern.1em\hat{\mathcal{A}}
\right)\;. \label{eq_H_rescaled_operators}
\end{equation}
Semiclassics concerns the limit in which the action is large compared to unity ($\hslash$ in conventional units). From \eqref{eq_H_rescaled_operators}, 
we see that this requires ${\overline n}$ to be large. For the limit to be meaningful, however, the expression in the brackets should take on an ${\overline n}$-independent value. 
Thus, $U\kern.1em\overline{n}$ needs to be held fixed. In summary, therefore, we find that the semiclassical limit for the Bose-Hubbard model amounts to
\begin{equation}
{\overline n} \rightarrow \infty \qquad \qquad U\kern.1em{\overline n} = \text{const.}
\label{eq_semiclasslim0DBH}
\end{equation}
In practice, the kind of situation one has in mind entails a cold bosonic gas which is dilute, and hence weakly interacting (i.e. small $U$), 
but comprises a large number of atoms, so that $U\kern.1em{\overline n}$ is a finite value comparable with $\omega_{\rm e}$.

\subsection{Herman-Kluk theory for 0D Bose-Hubbard model}
Having all the preliminaries in place, we now proceed to the application of Herman-Kluk theory to the 0D Bose-Hubbard model.
\vspace{-.75em}
\paragraph{Classical quantities} The first step is to calculate the classical ingredients that `feed' the HK propagator in \eqref{eq_HKgen}. 
The Hamilton function, which generates the underlying classical dynamics, is found using the usual substitution $\hat{a} \rightarrow z$ to yield
\begin{equation}
H(z,z^*) = \omega_{\rm e} |z|^2 + \tfrac{1}{2}U|z|^4\;.
\end{equation}
The equations of motion for the classical trajectory is hence, according to \eqref{eq_canonicaleom}:
\begin{equation}
{\rm i}\kern.05em\partial_t z = \partial_{z^*}H = \left(\omega_{\rm e} + U|z|^2\right)\kern-.1em z\;. \label{eq_eomconcrete}
\end{equation}
At first glance, this is a system of two coupled non-linear ordinary differential equations for the real and imaginary part of $z$ respectively, 
due to the explicit dependence of the right-hand side on both $z$ and $z^*$. However, we can exploit the underlying symmetry of the Hamiltonian, namely 
the $U(1)$ symmetry,---which, in the quantum mechanical version, led to conservation of $\hat{n}$---to observe that $|z|^2$ is a constant of 
motion,\footnote{A more pedestrian way of seeing this would be to explicitly calculate $\partial_t |z|^2 = \partial_t (z^* z)$ using product rule and 
to subsequently simplify using \eqref{eq_eomconcrete} and its complex conjugate.} i.e. $|z|^2 = |z_0|^2$. Thus, \eqref{eq_eomconcrete} effectively describes 
a harmonic oscillator, but with amplitude-dependent frequency; the latter being the manifestation of the anharmonicity in the classical picture. The solution is hence
\begin{equation}
z(t) \equiv z_t\kern-.025em\!\left(z_0^{\vphantom{*}},z_0^*\right) = \exp\!\left[-{\rm i}\kern.05em t\!\left(\omega_{\rm e} + U|z_0|^2\right)\right]z_0
\label{eq_solclasstraj}
\end{equation}
In order to evaluate the action $S_t\kern-.025em(z_0^{\vphantom{*}},z_0^*)$ along a classical trajectory $z_t\kern-.025em(z_0^{\vphantom{*}},z_0^*)$, one can 
insert the equation of motion \eqref{eq_eomconcrete} into the definition of the Lagrangian \eqref{eq_defS2} to obtain, for the specific problem at hand,
\begin{equation}
L_t\kern-.025em\!\left(z_0^{\vphantom{*}},z_0^*\right) = \tfrac{1}{2}U\!\left|\kern.01emz_t\kern-.025em\!\left(z_0^{\vphantom{*}},z_0^*\right)\kern-.05em\right|^{\kern.01em 4}\;,
\end{equation}
i.e. the Lagrangian along a classical trajectory is simply the anharmonic part of the Hamiltonian along that trajectory. Nevertheless, it turns out that the 
Lagrangian is conserved along a classical trajectory: Since $|z|^2 = |z_0|^2$, it follows that $|z|^4 \propto L_t(z_0^{\vphantom{*}},z_0^*)$ is also conserved. 
Thus, the action is easy to evaluate, and we find
\begin{equation}
S_t\kern-.025em\!\left(z_0^{\vphantom{*}},z_0^*\right) = \tfrac{1}{2}Ut\kern.05em|z_0|^4\;.
\label{eq_solclassaction}
\end{equation}
The final ingredient that enters the HK propagator is the prefactor. The stability factor is a straightforward computation of a derivative 
$\partial_{z_0} z_t$ to yield
\begin{equation}
R_t^{\rm \kern.05em HK}\kern-.05em\!\left(z_0^{\vphantom{*}},z_0^*\right) 
= [\partial_{z_0} z_t\kern-.025em\!\left(z_0^{\vphantom{*}},z_0^*\right)]^{1/2} 
= \sqrt{1 - {\rm i}\kern.05emU\kern-.05em t|z_0|^2} \, \exp\!\left[-\tfrac{1}{2}{\rm i}\kern.05em t\!\left(\omega_{\rm e} + U|z_0|^2\right)\right]
\label{eq_solclassprefactor}
\end{equation}
Note that the square root of the imaginary exponential is not the principal square root, but rather chosen such that the phase is continuous in time. Lastly, for the phase correction factor, we need the mixed second order partial derivative $\partial_{z^*} \partial_{z} H = \omega_{\rm e} + 2U|z|^2$, whence, due to conservation of $|z|^2$, we obtain
\begin{equation}
\theta_t\kern-.025em\!\left(z_0^{\vphantom{*}},z_0^*\right) = \left(\tfrac{1}{2}\omega_{\rm e} + U\kern-.05em |z_0|^2\right)\kern-.1emt\;.
\end{equation}
For the prefactor, we have
\begin{equation}
R_t\kern-.025em\!\left(z_0^{\vphantom{*}},z_0^*\right) = \sqrt{1 - {\rm i}\kern.05emU\kern-.05em t|z_0|^2} 
\, \exp\!\left(\tfrac{1}{2}{\rm i}\kern.1em Ut|z_0|^2\right)
\end{equation}

\vspace{-.75em}
\paragraph{Evaluating the HK propagator} With the classical ingredients at hand, all that remains is to put everything together according to the 
prescription \eqref{eq_HKgen}. 
The natural expression for the HK-SCIVR---for the purpose of general formalism---is as an integral over dyadic products of coherent states of the form $\ket{z_t\kern-.025em(z_0^{\vphantom{*}},z_0^*)}\!\bra{z_0}$.  However, for a subsequent comparison with the exact solution \eqref{eq_Uspectral}, which is given in spectral representation, the occupation number basis $\{\ket{n}\}$ is more suitable. In order to respect this, we directly consider occupation number matrix elements of the form 
$U_{nn^\prime}^{\rm HK}(t) \equiv \braket{n | \hat{U}_{\rm HK}(t) | n^\prime}$. Inserting the identity
\begin{equation}
\braket{n | z} = \frac{z^n}{\sqrt{n!}}\,{\rm e}^{-\frac{1}{2}|z|^2}
\end{equation}
for the necessary overlaps of occupation number eigenstates with coherent states, we obtain\footnote{Note that $|z|^2$ is conserved along a classical trajectory, so that the Gaussian weight arising from the normalization of the coherent states depends only on $|z_0|^2$.}
\begin{equation}
\begin{split}
U_{nn^\prime}^{\rm HK}(t) & = {\rm e}^{-{\rm i}n\omega_{\rm e}t}\!\varint{}{}\frac{{\rm d}^2 z_0}{\pi}\,{\rm e}^{-\left|z_0\kern-.075em\right|^2}\!\sqrt{1 - {\rm i}\kern.05emU\kern-.05em t|z_0|^2}\,\exp\!\left[{\rm i}\kern.1em Ut\kern.1em\Bigl(\tfrac{1}{2}|z_0|^4 - \left(n - \tfrac{1}{2}\right)\!|z_0|^2\Bigr)\right] \\ & \hphantom{= {\rm e}^{-{\rm i}n\omega_{\rm e}t}\!\varint{}{}\frac{{\rm d}^2 z_0}{\pi}}\!\!\!\! \times\frac{z_0^n (z_0^*)^{n^\prime}}{\sqrt{n!n^\prime!}}\;.
\end{split}
\end{equation}
Introducing new integration variables $z_0 = \sqrt{s}\kern.15em{\rm e}^{{\rm i}\varphi}$, this becomes
\begin{equation}
\begin{split}
U_{nn^\prime}^{\rm HK}(t) & = {\rm e}^{-{\rm i}n\omega_{\rm e}t} \varint{0}{\kern.1em \infty} \!{\rm d}s\,{\rm e}^{-s} \sqrt{1 - {\rm i}\kern.05emU\kern-.05em ts}\,\exp\!\left[{\rm i}\kern.1em Ut\kern.1em\Bigl(\tfrac{1}{2}s^2 - \left(n - \tfrac{1}{2}\right)\!s\Bigr)\right]\frac{s^{\frac{1}{2}(n + n^\prime)}}{\sqrt{n!n^\prime!}} \\ & \qquad\times \frac{1}{2\pi} \varint{0}{\kern.1em 2\pi}\!{\rm d}\varphi\,{\rm e}^{{\rm i}(n-n^\prime)\varphi}\;.
\end{split}
\end{equation}
The angular integration yields a Kronecker delta, thereby automatically ensuring that the HK-SCIVR respects the diagonal structure of the 
(exact quantum mechanical) propagator in the occupation number basis. We thus find
\begin{align}
U_{nn^\prime}^{\rm HK}(t) & = \delta_{nn^\prime}\,{\rm e}^{-{\rm i}n\omega_{\rm e}t}\, g_n(U\kern-.05emt)\;,
\label{eq_HK0DBH}
\end{align}
where we have introduced, for convenience of notation and future reference, the function $g_n(\cdot)$ by its integral representation
\begin{equation}
g_n(\tau) \equiv \frac{1}{n!}\varint{0}{\kern.1em\infty}\!{\rm d}s \,{\rm e}^{-s}s^n \sqrt{1 - {\rm i}\kern.05em\tau s}\,
\exp\!\left[{\rm i}\kern.1em \tau \kern.1em\Bigl(\tfrac{1}{2}s^2 - \left(n - \tfrac{1}{2}\right)s\Bigr)\right]\;.
\label{eq_defgn}
\end{equation}
Note that for $U = 0$, we have
\begin{align}
U_{nn^\prime}^{\rm HK}(t) & = \delta_{nn^\prime}\,{\rm e}^{-{\rm i}n\omega_{\rm e}t}\, g_n(0) = \delta_{nn^\prime}\,{\rm e}^{-{\rm i}n\omega_{\rm e}t}\, \frac{1}{n!}\varint{0}{\kern.1em\infty}\!{\rm d}s \,{\rm e}^{-s}s^n \nonumber \\
& = \delta_{nn^\prime}\,{\rm e}^{-{\rm i}n\omega_{\rm e}t}\, \frac{\Gamma(n+1)}{n!} = \delta_{nn^\prime}\,{\rm e}^{-{\rm i}n\omega_{\rm e}t}\;.
\end{align}
The \add{equidistant energy spectrum in the exponent} is of course the expected exact result for vanishing anharmonicity, which may serve as a first 
consistency check. The nontrivial features manifest themselves at nonzero interaction strength, to which we turn our attention in the following.

\section{Analytical results}
\label{sec_anal_results}
Having established the HK propagator for the 0D Bose-Hubbard model in the previous section, the purpose of the remainder of this investigation is 
to study it and see what understanding may be gleaned from it. The primary question on which we shall concentrate is how well it agrees with the exact 
quantum mechanical solution, which we already have as a convenient closed-form expression. Since we have an explicit integral representation, it is 
reasonable to first attempt an approximative evaluation of the integral by analytical means, especially under sufficiently nontrivial limiting conditions.

To this end, we consider now the limit $n \rightarrow \infty$ with $\tilde{\tau} \equiv n\tau = O(1)$ fixed and make the variable substitution $s = n\kern.05em x$ 
in the integral \eqref{eq_defgn}. This is chosen so as to mimic the semiclassical limit in the sense of eq. \eqref{eq_semiclasslim0DBH}. The first condition 
is the limit of `large quantum numbers' familiar from, e.g., Bohr-Sommerfeld quantization and similar semiclassical schemes. To see that the rescaling 
effected through the change of integration variables is indeed appropriate---especially in view of \eqref{eq_rescaledops}---, note that the correspondence 
rule is $\hat{a} \rightarrow z$, and that $s = |z|^2$. We now have
\begin{equation}
g_n(\tau) = \frac{n^{n+1}}{n!}\varint{0}{\kern.1em\infty} \! {\rm d}x\,f(x)\kern.1em{\rm e}^{n\tilde{S}(x)}\;,
\end{equation}
where 
\begin{align}
f(x) = {\rm e}^{\frac{1}{2}{\rm i}\tilde{\tau}x}\sqrt{1 - {\rm i}\tilde{\tau}x} \quad \text{and} \quad \tilde{S}(x) = \ln x - x + {\rm i}\tilde{\tau} \left(\tfrac{1}{2}x^2 - x\right)\!\;.
\end{align}
This lends itself to a steepest descent argument \add{\cite{lit_MathewsWalker}}. {In the limit $n\to\infty$, the critical points of $\tilde{S}$ alone, which are given by}
\begin{equation}
x_0 = 1 \qquad \text{and} \qquad x_1 = ({\rm i}\tilde{\tau})^{-1}.
\end{equation}
{have to be considered.} Since $f(x_1) = 0$, the second stationary point has no contribution, and we can proceed as though there were only one stationary point. We first evaluate, 
using elementary calculus and some algebra, the quantities required for the method of steepest descent:
\begin{align}
f(x_0) & = {{\rm e}^{\frac{1}{2}{\rm i}\tilde{\tau}}}\sqrt{1 - {\rm i}\tilde{\tau}}\;, \nonumber\\
\tilde{S}(x_0) & = -\bigl(1 + \tfrac{1}{2}{\rm i}\tilde{\tau}\bigr)\;, \\
(\partial_x)^2 \tilde{S}\!\left.\vphantom{S^a_j}\right|_{x=x_0} & = -(1 - {\rm i}\tilde{\tau})\;. \nonumber
\end{align}
The steepest descent approximation then yields
\begin{equation}
g_n(\tau) = \frac{n^{n+1}}{n!}\sqrt{\frac{2\pi}{n}}\frac{{\rm e}^{n\tilde{S}(x_0)}}{\sqrt{-(\partial_x)^2 \tilde{S}\!\vphantom{S^a_j}\kern.2em\big|_{x=x_0}}}\bigl(f(x_0) + O(1/n)\bigr)\;.
\end{equation}
Inserting the ingredients calculated beforehand and recalling that $\tilde{\tau} \equiv n\tau$, the resulting asymptotic behaviour is
\begin{equation}
g_n(\tau) \simeq \frac{\sqrt{2\pi n}\,(n/e)^n}{n!}\,{\rm e}^{-\frac{1}{2}{\rm i}\tau n(n-1)} \simeq {\rm e}^{-\frac{1}{2}{\rm i}\tau n(n-1)}\;,
\end{equation}
where the second asymptotic relation is valid due to Stirling's formula. Putting $\tau = U\kern-.05emt$ according to \eqref{eq_HK0DBH} and comparing with 
the spectral representation \eqref{eq_Uspectral}, one finds the main result of this section: In the limit $n \rightarrow \infty$, $Un = \text{const.}$, 
the spectrum of the HK propagator is
\begin{equation}
E_n^{\rm HK} \simeq \omega_{\rm e}n + \tfrac{1}{2}Un(n-1)\;,
\end{equation}
which is in \add{full} agreement with the exact result \add{(\ref{eq_exactspectrum})}.

In the following, we discuss the significance of two particular terms that are in some sense a `special feature' of the HK propagator, viz. the phase correction 
$\theta_t$ and the stability factor $R_t$, by examining what would happen if one were to neglect either of these contributions. 
The role played by the phase correction in ensuring the correct energy spectrum for the harmonic oscillator has already been remarked upon---a global factor 
comes about and accounts for the ground-state energy. Here, however, $\theta_t$ contributes a `local' factor as a function of the initial value $z_0$ of the 
underlying classical trajectory, and is therefore, in some sense, more involved. Indeed, only after evaluating the resultant phase space integral 
in steepest descent approximation---i.e., physically, making the transition to the semiclassical limit---does its role become clear. Note that carrying out 
essentially the same steps as above, but with $\theta_t = 0$, yields a similar result, but with
\begin{equation}
\left.E_n^{\rm HK}\right|_{\theta_t = 0} \simeq \omega_{\rm e}\!\left(n + \tfrac{1}{2}\right) + \tfrac{1}{2}Un(n + 1)\;.
\end{equation}
The universal feature, it would appear, is that the leading order (LO) behaviour of the energy, viz. $\omega_{\rm e}n\kern.1em ({1} + O(1/n))$ and 
$\tfrac{1}{2}Un^2(1 + O(1/n))$ for the harmonic and quartic parts of the Hamiltonian respectively, is captured correctly. The phase correction factor is 
important to ensure that the next-to-leading order (NLO) term---which happens to be the ground-state energy in the harmonic case and linear in $n$ for the 
quartic part---is also reproduced exactly.

As a final analytical demonstration, we investigate what would happen if one were to work with the so-called Frozen Gaussian Approximation (FGA). 
In generic problems, the trajectories are not available analytically, so that one is forced to integrate equations of motion for the monodromy matrix, in addition 
to the trajectory (cf. e.g. \cite{lit_FemtosecPhys}, or \cite{lit_Child} for the complex version). 
An often used method, especially in theoretical chemistry, is to simply neglect the time evolution of $\underline{{\bf M}}$, i.e. to put 
$\underline{{\bf M}}(t) = \underline{{\bf M}}(0) = 1\kern-.3em1$ (cf. e.g. \cite{lit_TatchenPollak}). 
This amounts to setting $R_t^{\rm \kern.05em HK} = 1$ in the above. Redoing all the steps in an analogous manner, one finds in the semiclassical limit
\begin{equation}
U_{nn^\prime}^{\rm FGA}(t) \simeq \delta_{nn^\prime}\, {\rm e}^{-{\rm i}\left(n-\frac{1}{2}\right)\omega_{\rm e}t}\,\frac{{\rm e}^{-\frac{1}{2}{\rm i}U\kern-.05em n(n-2)t}}{\sqrt{1 - {\rm i}\kern.1em n\kern.05em U\kern-.05em t}}\;.
\label{eq_FGAsemiclasslim}
\end{equation}
For both the harmonic and anharmonic parts of the Hamiltonian, the FGA yields the correct LO dependency, but results in an erroneous 
NLO behaviour---just as we observed for the phase correction factor. It is well-known from literature that the ground-state energy is reproduced wrongly 
by the FGA. This is also apparent from \eqref{eq_FGAsemiclasslim}. Something similar occurs for the anharmonic part, where the 
contribution linear in $n$ has the wrong coefficient.  The disagreement in the anharmonic case, however, goes further than just the energy spectrum: 
The exact (and, in the semiclassical limit, the HK as well \cite{lit_Herman}) propagator 
is unitary, i.e. $\hat{U}^\dagger\kern-.1em(t)\kern.1em\hat{U}(t) = 1\kern-.3em1$. Expressed in terms of matrix elements, it follows that 
\[|U_{nn^\prime}(t)|^2 = \braket{n | \hat{U}^\dagger\kern-.1em(t)\kern.1em\hat{U}(t) | n^\prime} = \braket{n | n^\prime} = \delta_{nn^\prime}\;.\] For the 
FGA, however, we have directly from \eqref{eq_FGAsemiclasslim}
\begin{equation}
\left|U_{nn^\prime}^{\rm FGA}(t)\right|^2 = \frac{\delta_{nn^\prime}}{1 + n^2 U^2 t^2}\;.
\end{equation}
The faster loss of unitarity in the FGA has been attested to in the literature by means of numerical investigations \cite{lit_TatchenPollak}. The above 
considerations furnish an explicit demonstration of this effect. Furthermore, we see that the time-scale for the loss of unitarity (e.g. the `half-life' of 
diagonal matrix elements of $\hat{U}^\dagger\hat{U}$ in occupation number basis) is $\propto 1/U$, i.e. the higher the anharmonicity of the system, the more 
rapidly unitarity is lost.

\section{Numerical results}
\label{sec_numresults}
The only nontrivial part of the expression \eqref{eq_HK0DBH} for the HK propagator is the function $g_n(\tau)$. In order to analyse this quantity for 
generic parameter values, especially beyond the semiclassical limit, we employ numerical integration. The highly oscillatory nature of the integrand, 
which afforded scope for the application of steepest descent asymptotics in Section \ref{sec_anal_results}, is now the primary difficulty one faces when 
attempting a straightforward numerical evaluation of the integral. In order to generate the numerical results we shall discuss in the following, a global 
adaptive Gauß-Konrod quadrature was employed, using the built-in implementation \texttt{NIntegrate} of the \texttt{Mathematica} language.\footnote{The higher 
the value of $n$, the quicker the integrand oscillates. The sign problem ensuing thence was solved in a brute-force manner, by increasing 
the working precision so that it exceeds the desired precision by a sufficient margin. Thus, for $n \sim 30$, the intermediate calculations had to be carried 
out to $\sim 100$ decimal places in order to obtain a final accuracy of $5$ decimal places.} We consider a modified polar representation of the function $g_n(\tau)$, i.e.\begin{equation}
g_n(\tau) = \left|g_n(\tau)\right|\kern.025em {\rm e}^{-{\rm i}\varphi_n(\tau)}\;,
\end{equation}
where $\varphi_n$ is chosen to be continuous\footnote{A proper polar representation would require $\varphi_n(\tau) \in \left[-\pi,\pi\right)$ for all $\tau$} 
in $\tau$. The decomposition into the above form is opportune because the absolute value and the phase correspond to two physically relevant properties: conservation of unitarity and the energy spectrum.

\subsection{Conservation of unitarity}
\begin{figure}
\centering
\includegraphics[width=0.6\textwidth]{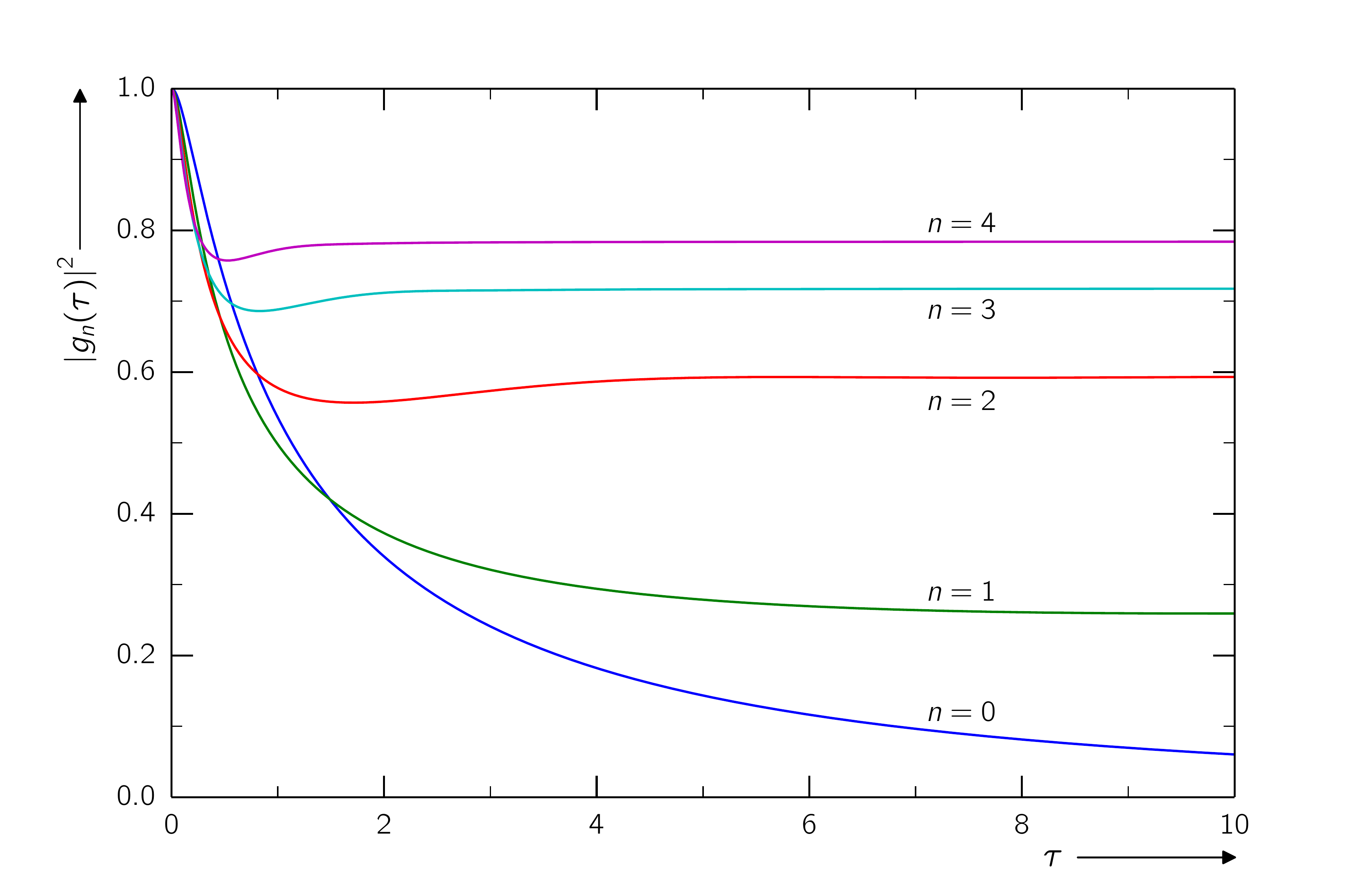}
\caption{Loss of unitarity for low values of $n$, as indicated by the decay of the absolute value $\left|g_n(\tau)\right|^2$, which represents the norm of 
the propagator in the sense of \eqref{eq_connection_norm}.}
\label{fig_r_n_lown}
\end{figure}

\begin{figure}
\centering
\includegraphics[width=0.6\textwidth]{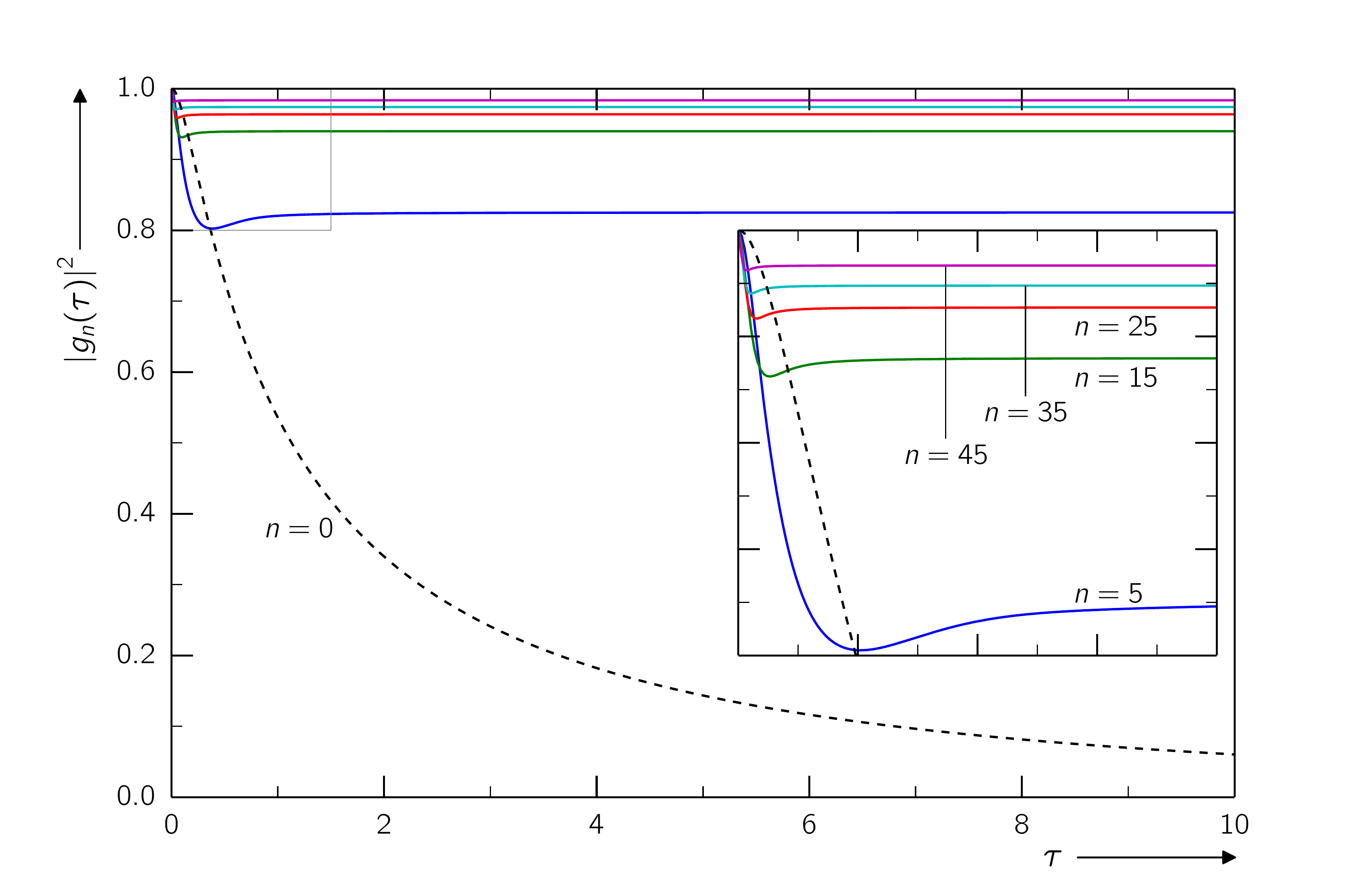}
\caption{Recovering unitarity for large values of $n$, as evidenced by the behaviour of the norm $\left|g_n(\tau)\right|^2$. The dotted line shows $n = 0$ for 
comparison.}
\label{fig_r_n_highn}
\end{figure}

The quantity $|g_n\kern-.05em|$ is closely related to the conservation of unitarity, which is one of the basic properties of the (exact) time evolution operator. 
To see this more explicitly, observe that
\begin{equation}
\left|g_n(U t)\right|^{2} = \left|U_{nn}^{\rm HK}(t)\right|^{2} = \bra{n}\! \left(\hat{U}_{\rm HK}^\dagger\kern-.1em(t)\kern.1em\hat{U}^{\vphantom{\dagger}}_{\rm HK}\kern-.1em(t)\right)\! \ket{n}\;.
\label{eq_connection_norm}
\end{equation}
Alternatively, one may interpret $\left|g_n(U t)\right|^{2}$ as the norm of the wave function $\hat{U}_{\rm HK}(t)\!\ket{n}$. In Section \ref{sec_anal_results}, 
we established that $\left|g_n\right| \simeq 1$ for large quantum numbers, thus ensuring conservation of unitarity in the semiclassical limit. Fig.\ 
\ref{fig_r_n_lown} depicts numerical results for $0 \leqslant n \leqslant 5$. Especially for $n = 0$, which is evidently outside the scope of the semiclassical 
regime, the norm is seen to decay over time and go to zero. For higher values of $n$, the norm still decays for a short time but then stabilizes to a positive 
value between zero and unity. The time scale on which the decay and stabilization of the norm takes place becomes smaller for increasing $n$, and the stable 
value approached by the norm increases with $n$ and approaches unity.
This interpretation is reinforced by Fig. \ref{fig_r_n_highn}, which shows the behaviour of $\left|g_n\right|^2$ for higher values of $n$, where the role of semiclassics is more pronounced. Apart from a very short (negligible) interval where the decay and stabilization takes place, one may effectively think of it as a step function, to wit:
\begin{equation}
\left|g_n(\tau)\right| \approx \begin{cases} 1 & \tau = 0 \\ r_n & \tau > 0 \end{cases}
\end{equation}
with $r_{n} \rightarrow 1$ for $n \rightarrow \infty$. This furnishes a numerical verification that one does, indeed, obtain $\left|g_n\right| \simeq 1$ for 
large quantum numbers, thus conserving unitarity in the semiclassical limit.

\subsection{Energy spectrum}
\begin{figure}
\centering
\includegraphics[width=0.6\textwidth]{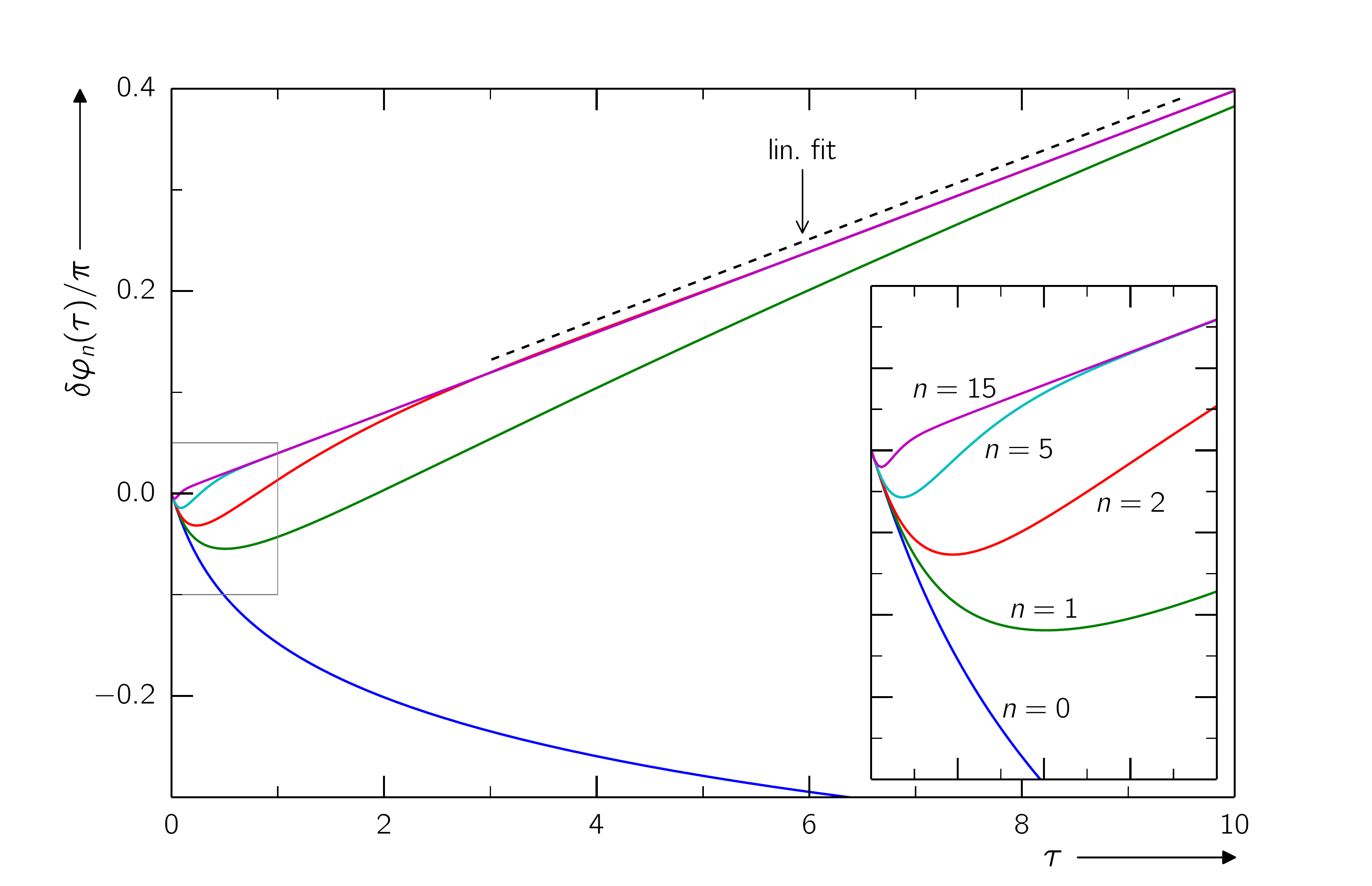}
\caption{Plot of $\delta\varphi_n(\tau)$. The limiting behaviour for large quantum numbers is well approximated by the linear fit 
$\delta\varphi_n(\tau) \simeq 0.125\tau$, which is depicted (black dashed line) with a small random offset for reasons of visibility.}
\label{fig_phi_n}
\end{figure}

The phase $\varphi_n(\tau)$ is closely related to the spectrum: If $\varphi_n$ is linear in $\tau$, then
\begin{equation}
\varphi_n(Ut) = E_n^{\rm anh}t\;,
\end{equation}
with $E_n^{\rm anh} = E_n - \omega_{\rm e}n$ the anharmonic part of the energy. We write
\begin{equation}
\varphi_n(\tau) = \tfrac{1}{2}n(n-1)\tau + \delta\varphi_n(\tau)\;,
\end{equation}
where we have anticipated that {result based on} the HK propagator should deviate from the exact solution by some small amount $\delta\varphi_n(\tau)$. This is plotted in 
Fig. \ref{fig_phi_n}. For small quantum numbers, as was the case with unitarity, there is a significant deviation from the exact solution---which would be a 
constant line at $\delta\varphi_n = 0$. This deviation does not, however, vanish for higher values of $n$. Instead, the limiting behaviour for large quantum 
numbers was found (by linear fitting) to be 
\begin{equation}
\delta\varphi_n(\tau) \simeq 0.125\kern.1em\tau\;.
\end{equation}
{
This observation motivated us to revisit the steepest descent calculations of Section \ref{sec_anal_results}.
The function $f(x)$ contributes a phase term which is NLO in $1/n$ and linear in $x$. Including this term in the determination of the stationarity condition, the first (dominant) critical point in the limit $\tilde{\tau}\gg 1$ becomes $x_0=1-1/2n$. Performing the same steps as in Section \ref{sec_anal_results} with this
modified critical point, we obtain
\begin{equation}
 g_n(\tau) = {\rm e}^{-\frac{1}{2}{\rm i}\tau n(n-1)-\frac{1}{8}{\rm i}\tau}\;,
\end{equation}
that is, there is a next-to-next-to-leading order (NNLO) contribution to the spectrum, exactly identical to the one that we found numerically.
\void{
The question, then, is whether this is a small discrepancy---and in what sense, if that be the case. As mentioned previously, the phase function 
$\varphi_n(\tau)$ is closely related to the energy spectrum. Since the relation is linear in $\tau$, we can immediately read off the error in 
energy resulting from the deviation described above. Recalling that the HK propagator is exact for the harmonic part of the Hamiltonian, we see that the error arising from $\delta \varphi_n$, i.e. the anharmonic part, directly gives the total error of the HK spectrum. With $\delta\varphi_n(Ut) = 0.125\kern.1em U\kern-.05emt$, we thus obtain
\begin{equation}
\delta E_n^{\kern.05em\rm HK} = 0.125\kern.1em U\;.
\end{equation}
The relative error, hence, is found to be
\begin{equation}
\frac{\delta E_n^{\kern.05em \rm HK}}{E_n} \propto \frac{1}{n^2\bigl(1 + O(1/n)\bigr)} = O\kern.05em\bigl(1/n^2\bigr)\;.
\end{equation}
}
Thus, the HK propagator is not entirely exact when predicting the energy levels of the quartic part of the Bose-Hubbard Hamiltonian. 
Although in the semiclassical limit ($n\to \infty$), it reproduces all energies accurately, it incurs an error in the NNLO term
if the $x$ integration is performed either numerically or with finite $n$ corrections to the steepest descent approximation.} 
\void{
Note that the analytical 
treatment using steepest descent was unable to anticipate this error, since the decay $\propto 1/n^2$ is `too fast'. The question whether a higher order steepest 
descent, e.g. using Watson's lemma in conjunction with appropriate variable substitutions, would have been able to predict the N\textsuperscript{2}LO error 
is left as an open problem.
}

\section{Wave packet dynamics}
\label{sec_wavepacketdyn}

\begin{figure}[b!]
  \centering
   \includegraphics[width=.4\textwidth, angle = 0]{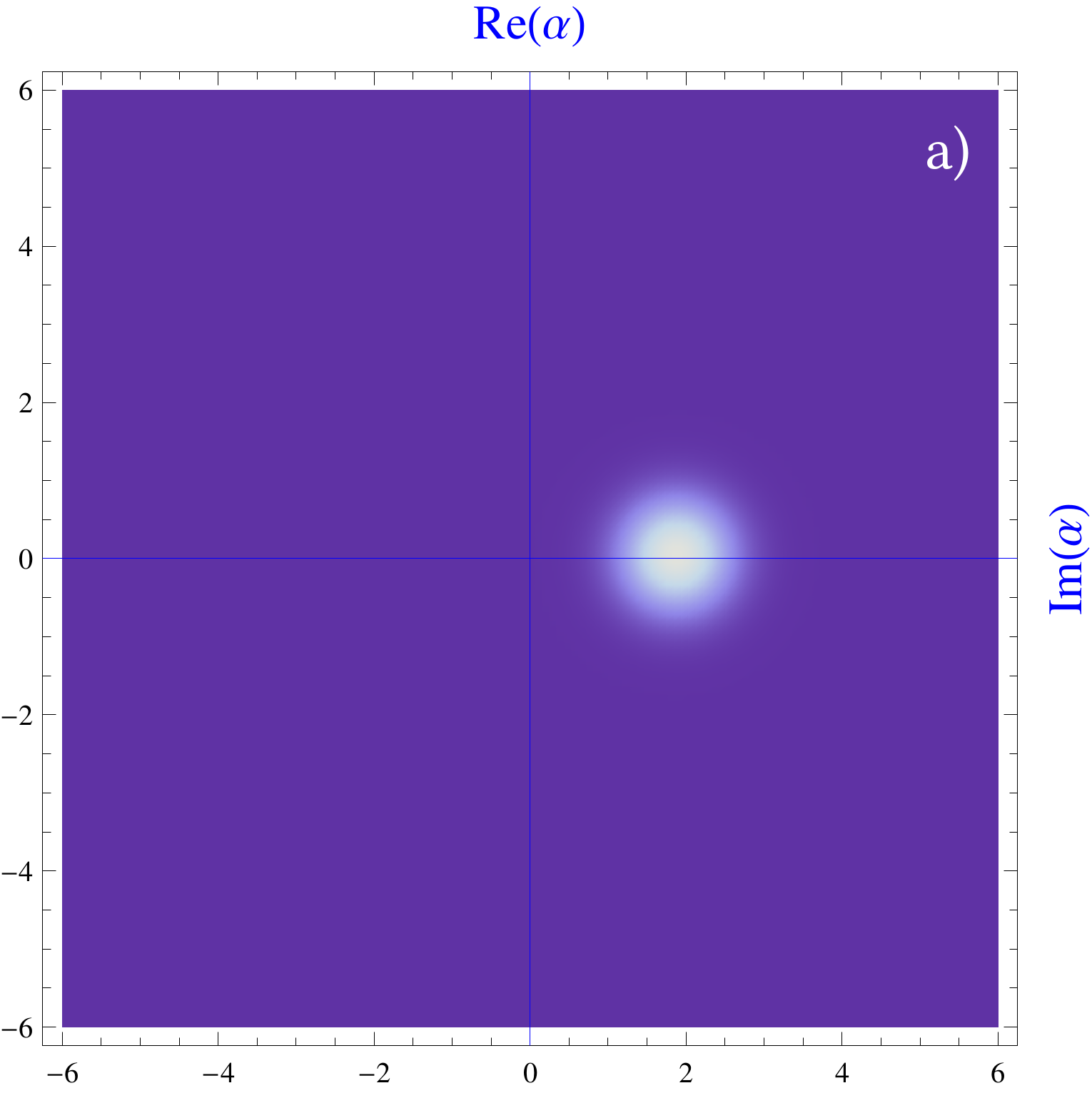}\hspace*{.5em}
   \includegraphics[width=.4\textwidth, angle = 0]{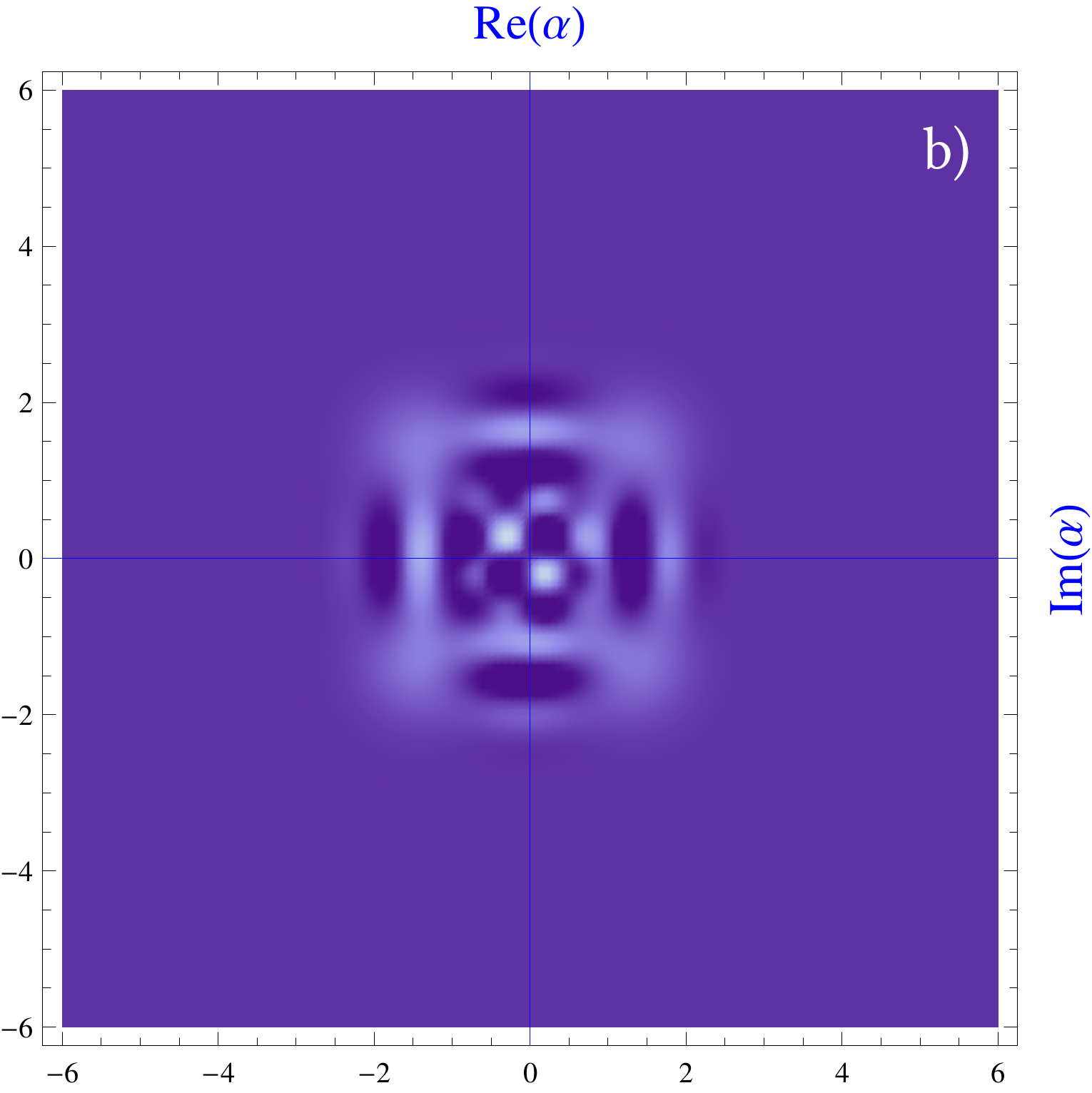}\\[.5em]
   \includegraphics[width=.4\textwidth, angle = 0]{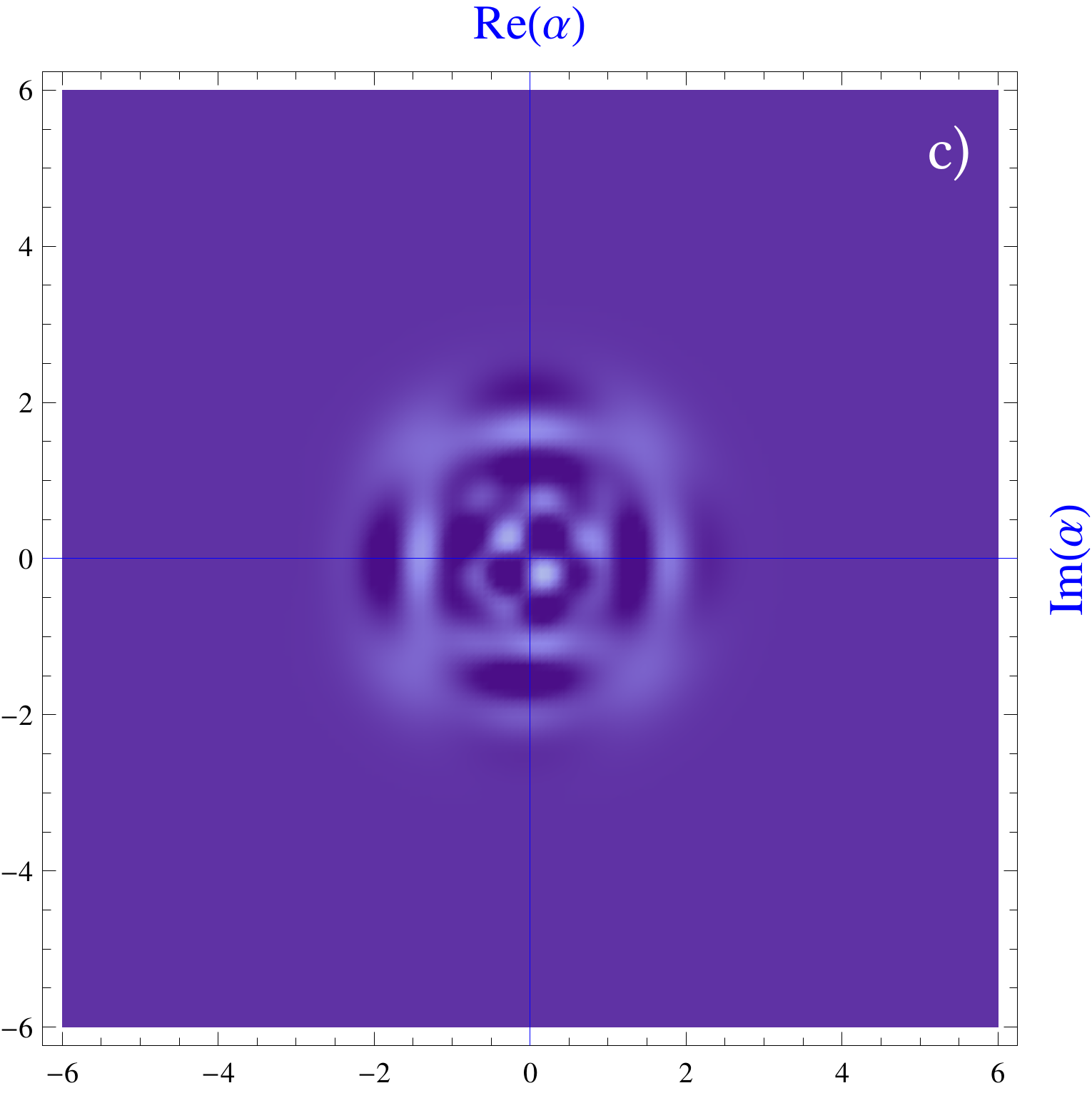}\hspace*{.5em}
   \includegraphics[width=.4\textwidth, angle = 0]{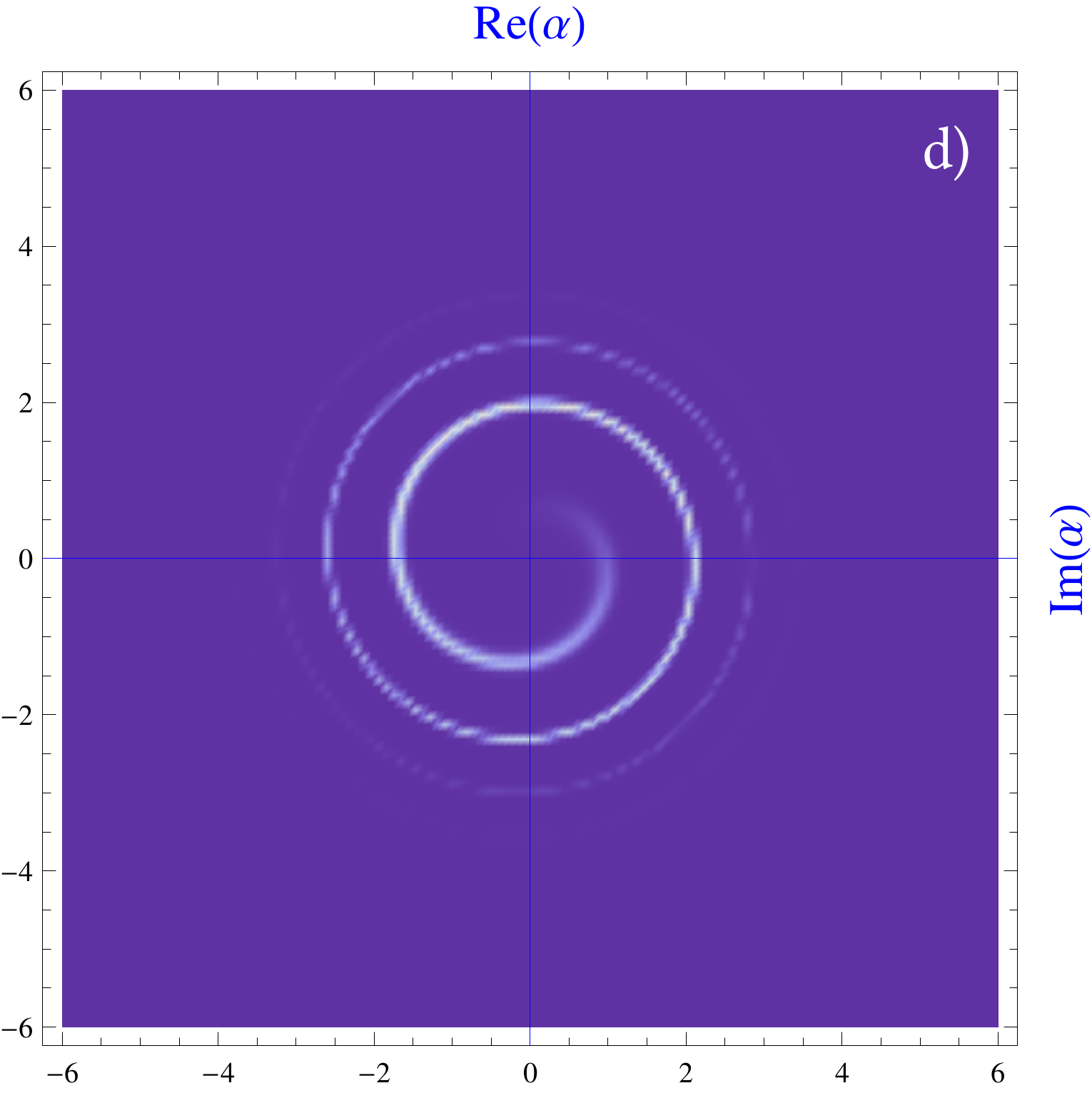}\\[1em]
   \includegraphics[scale=0.5]{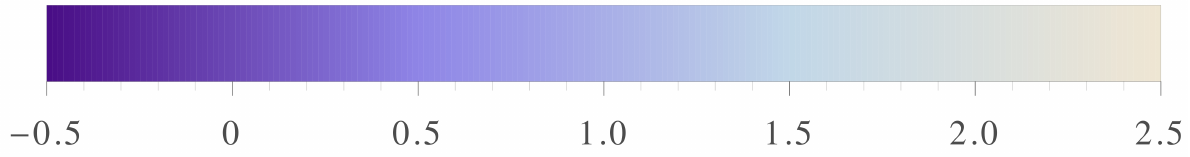}
  \caption{Wigner function $W(\alpha,\alpha^*,t)$; a) initial coherent state, b)-d) Wigner function after a time 
  $\omega_{\rm e} t=5\cdot 2\pi$ and with $U = 0.05$. In b) exact Wigner function, c) Herman-Kluk IV representation, 
  d) Truncated Wigner approximation.}
 \label{Fig_Wig}
\end{figure}

The quality of the Herman-Kluk IVR is further elucidated by investigating the dynamics of an initial wave packet with non-linear Hamiltonian \eqref{eq_0DBHHamiltonian}. The corresponding quantum state
is visualized using the Wigner function \cite{lit_quantum_optics_book}, in complex quantum optical notation
\begin{equation}
 W(\alpha,\alpha^*) = \varint{}{}\frac{{\rm d}^2\eta}{\pi}\;{\rm e}^{\eta^*\alpha-\eta\alpha^*}\! \chi(\eta,\eta^*)\;,
 \label{eq_wigner}
\end{equation}
where $\alpha=(q+ip)/\!\sqrt{2}$ is the phase space variable. Here, the complex Fourier transform is performed over phase space ${\rm d}^2\eta={\rm d}(\operatorname{Re}\eta)\kern.1em {\rm d}(\operatorname{Im}\eta)$, and
the characteristic function is determined from the underlying quantum state $\hat{\varrho}$ as the expectation 
value of the displacement operator:
\begin{equation}
 \chi(\eta,\eta^*)=\operatorname{Tr}\kern-.15em\left(\hat{\varrho}\, {\rm e}^{\eta \hat{a}^\dagger - \eta^* \hat{a}}\right)\!\;.
\end{equation}
Throughout this section we choose a coherent initial state  $\ket{\psi(0)} = \ket{z_{\rm i}}$,
whose Wigner function is a simple Gaussian (see Fig. \ref{Fig_Wig}a) centered 
at $\alpha\approx z_{\rm i}=2$:
\begin{equation}
 W_0(\alpha,\alpha^*)=2\kern.1em {\rm e}^{-2|\alpha-z_{\rm i}|^2}\,.
\end{equation}
Due to the underlying non-linear dynamics, the Wigner function will not remain of Gaussian shape. 
Still, in our case there is an analytical exact expression based on an expansion of the wave packet in terms of number states \cite{Groenewold}, leading to
\begin{equation}
\begin{split}
 W(\alpha,\alpha^*,t)=2\kern.1em {\rm e}^{-|z_0|^2-2|\alpha|^2} & \Bigg\{
 \sum_{n=0}^\infty \frac{(-1)^n |z_0|^{2n}}{n!} L_n^0\bigl(4|\alpha|^2\bigr) + {\color{white}blindtext} \\ & \hphantom{\Bigg\{}\sum_{n=1}^\infty
 \sum_{m=0}^{n-1}\frac{(-1)^m}{n!}L_n^{n-m}\bigl(4|\alpha|^2\bigr) \!\left({\rm e}^{-{\rm i}(E_n-E_m)\kern.05em t}z_0^n\!\left(z_0^*\right)^{m}\!(2\alpha^*)^{n-m} +\text{c.c.}\right)\!\Bigg\}\;.
 \nonumber
\end{split}
\end{equation}
In this expression the eigenenergies are $E_n=\omega_{\rm e} n +\frac{1}{2}Un(n-1)$ from eq. \eqref{eq_exactspectrum}  and the
$L_n^k(x)$ are associated Laguerre polynomials \cite{Abramowitz_Stegun}. In Fig. \ref{Fig_Wig}b), we show
this exact Wigner function for a time evolution of duration $\omega_{\rm e} t = 5\cdot 2\pi$ and $U=0.05$.

The Wigner function can easily be obtained from the Herman-Kluk initial value representation \eqref{eq_HKgen}
of the propagator. All ingredients are known analytically from eqs. \eqref{eq_solclasstraj}, \eqref{eq_solclassaction} and \eqref{eq_solclassprefactor}. The integrals over $z_0$ in \eqref{eq_HKgen} and
over $\eta$ in \eqref{eq_wigner} are performed numerically here. Most remarkably, despite the rather lowly
excited initial coherent state with $z_{\rm i}=2$ and the nonlinearity of the dynamics, the time evolved 
HK Wigner function displays an overall very good agreement with the exact result, as can be seen in
Fig. \ref{Fig_Wig}c). Note that the depicted results are obtained without recourse to any further (ad-hoc) renormalization, relying instead on the approximate conservation of unitarity implemented a priori in the formulation of the HK propagator \eqref{eq_HKgen}.

The dynamics of ultracold interacting Bose gases is often described with the help of the so-called 
{\it truncated Wigner approximation} (TWA), where the exact dynamics is replaced by the classical dynamics
of the field (the Gross-Pitaevskii equation). Indeed, expressing the exact von Neumann equation for 
the density operator $\hat{\varrho}$ and Hamiltonian \eqref{eq_0DBHHamiltonian} in terms of the Wigner function, one arrives at a partial differential equation
\begin{equation}
 \partial_t W(\alpha,\alpha^*,t) = \left[{\rm i}\kern-.15em\left(\omega_{\rm e}+U|\alpha|^2\right)\!
 \left(\alpha\kern.05em\partial_\alpha-\alpha^*\partial_{\alpha^*}\kern-.1em\right)
 -\tfrac{1}{4}{\rm i}\kern.1em U\!\left(\alpha\!\left(\partial_\alpha\right)^{2}\kern-.1em\partial_{\alpha^*} -
 \alpha^* \partial_\alpha\!\left(\partial_{\alpha^*}\kern-.1em\right)^{2}\right)\right]W(\alpha,\alpha^*,t)\;.\nonumber
\end{equation}
Taking into account first order derivatives only (TWA), we arrive at a classical Liouville equation for a
phase space density $W(\alpha,\alpha^*,t)$ with Hamilton's equation of motion
\begin{equation}
 {\rm i}\kern.05em\partial_t \alpha = \left((\omega_{\rm e}-U)+U|\alpha|^2\right)\kern-.1em\alpha\;.
\end{equation}
Since the solutions $\alpha(t)$ for the classical trajectories are known from eq. \eqref{eq_solclasstraj}, the 
Wigner function in TWA is easily expressed in terms of the initial Wigner function as
\begin{equation}
 W_{\kern.1em\rm TWA}(\alpha,\alpha^*,t) = 
 W_0\!\left(\alpha\kern.05em {\rm e}^{{\rm i}\kern.025em\left((\omega_{\rm e}-U)+U|\alpha|^2\right)\kern.05em t},\kern.05em \alpha^* e^{-{\rm i}\kern.025em\left((\omega_{\rm e}-U)+U|\alpha|^2\right)\kern.05em t}\right)\,, \label{twa}
\end{equation}
shown in Fig. \ref{Fig_Wig}d) for our initial coherent state. Since the latter is positive, so is its time evolved truncated Wigner 
approximation (\ref{twa}). Indeed, as can be seen in Fig. \ref{Fig_Wig}, while the region of large
$W(\alpha,\alpha^*)$ is captured well by the TWA, any interferences (and thus negative values of $W$)
present in the exact Wigner function cannot be reproduced by $W_{\kern.1em \rm TWA}$. Evidently, the Wigner function
obtained from the semiclassical HK propagator---despite also containing classical ingredients only---provides a much better approximation.  We mention in passing that the difference in quality between HK and TWA is much less pronounced for shorter times, i.e. as long as the wave packet is still more or less localized in phase space.

\section{Conclusion\add{s and Outlook}}
\label{sec_conclu}
In this paper, we investigated a single-mode interacting Bose system using the Herman-Kluk semiclassical initial value representation (HK-SCIVR). 
First, the underlying classical dynamics was solved analytically by exploiting the symmetries of the problem. Subsequently, the classical ingredients were put 
together and an explicit expression was found for the HK propagator in terms of an integral representation. The semiclassical limit was obtained by evaluating the 
aforementioned integral in steepest descent approximation following opportune variable transformations. In addition, the often-used Frozen Gaussian 
Approximation---which is essentially the HK propagator, but where the time evolution of the monodromy matrix is neglected---was treated in similar fashion for 
comparison. Finally, the integral was evaluated numerically in order to supplement the analytical considerations. Last but not least, having studied the properties of the propagator per se, we turned our attention to an example application, viz. wave packet dynamics. To this end, we computed the time evolution of the Wigner function using the HK-SCIVR and compared the results with the exact solution (available analytically) as well as the popular truncated Wigner approximation (TWA).

The results may be summarized as follows: (i) Unitarity is conserved in the semiclassical limit by the HK propagator. For smaller quantum numbers, however, unitarity is seen to decay over time. The FGA, on the other hand, was found to explicitly violate unitarity, even in the semiclassical limit. (ii) The HK propagator reproduces the energy spectrum to next-to-leading order (with small parameter $1/n$) in the limit of large quantum numbers, and an error occurs only in next order. In contrast, the FGA is only accurate to LO, and it was shown explicitly that an error is incurred in NLO. Philosophically, this is the classical trade-off between efficiency and accuracy: On one hand, a speed-up may be achieved by circumventing the computation of the time evolution of the 
monodromy matrix. On the other hand, one has to contend with an inferior approximation of the energy levels and loss of unitarity---the latter more likely the 
graver issue in practice. (iii) As to the description of wave packet dynamics, the classical TWA fails to capture any interference effects (and thus negative values of the Wigner quasiprobability distribution) present in the exact solution. By contrast, although still relying on classical information only, the HK-SCIVR is able to reproduce the salient features of the exact solution correctly.

These investigations pave the way for a semiclassical treatment of a truly multiple-site Bose-Hubbard chain.
Certain issues regarding our 0D BH model may be deemed worthy of a closer look. One may wish, for instance, to comprehend and demonstrate 
explicitly the initial decay and subsequent stabilization of the norm, which we observed only numerically. \void{In similar vein, it may be worthy of interest to 
attempt an analytical approach to the N\textsuperscript{2}LO term of the energy spectrum of the HK propagator.} {This issue appears} to hinge 
on the evaluation of the steepest descent approximation to higher orders, which is a nontrivial mathematical task. It may also be worthwhile to investigate the possibility of 
obtaining the correct higher-order behaviour using semiclassical means, e.g. by employing higher-order corrections to the propagator, such as the semiclassical asymptotic expansion of Kay \cite{lit_Kay} or Zhang and Pollak \cite{lit_ZhangPollak}. In such cases, it may also be instructive to include more involved terms 
in the Hamiltonian, for instance a general power term of the form $\hat{n}(\hat{n} - 1)(\hat{n} - 2)\cdots(\hat{n} - \nu)$, whose energy spectrum will then have 
non-vanishing coefficients up to the $\text{N}^{\nu}\text{LO}$ term.

{Furthermore, a natural question to ask is if also fermionic Hubbard models could be treated in a similar vein.
In this regard it is interesting to note that a semiclassical propagator in fermionic Fock space has recently been established \cite{Eetal14}.
Although this work is based on a van Vleck-Gutzwiller type of propagator, applications of time-dependent semiclassical many-body methodology using an SCIVR approach in the same vein as Herman and Kluk
may be envisioned.}

\titleformat{\section}{\normalfont\bfseries\normalsize}{\thesection.}{0.5em}{#1}
\section*{Acknowledgements}
W.\,T.\,S is grateful for enlightening discussions with Juan Diego Urbina. {P.\,O. and L.\,S.  acknowledge
support from the International Max Planck Research School (IMPRS),
Dresden, Germany. The authors would like to thank an anonymous referee
for a very helpful comment regarding the steepest descent calculations.}

\vspace*{.25em}
\titleformat{\section}{\normalfont\bfseries\normalsize}{\thesection.}{0.5em}{#1}
\setlength\bibhang{3em}
\setlength\bibsep{0em}

\end{document}